# Solar-Sail Trajectory Design for Multiple Near-Earth Asteroid Exploration Based on Deep Neural Networks

Yu Song[*], and Shengping Gong[†]

*Tsinghua University, 100084 Beijing, People's Republic of China*

**Abstract:** In the preliminary trajectory design of the multi-target rendezvous problem, a model that can quickly estimate the cost of the orbital transfer is essential. The estimation of the transfer time using solar sails between two arbitrary orbits is difficult and usually requires to solve an optimal control problem. Inspired by the successful applications of the deep neural networks in nonlinear regression, this work explores the possibility and effectiveness of mapping the transfer time for solar sails from the orbital characteristics using the deep neural networks. Furthermore, the Monte Carlo Tree Search method is investigated and used to search the optimal sequence considering a multi-asteroid exploration problem. The obtained sequences from preliminary design will be solved and verified by sequentially solving the optimal control problem. Two examples of different application backgrounds validate the effectiveness of the proposed approach.

## I. Introduction

The Near-Earth Asteroids (NEA) have become of increasing interest since the 1980s because of the scientific significance and greater awareness of the potential threat to the Earth. The NEAs are mostly unchanged in composition since the early days of the solar system [1], and it is now widely accepted that collisions in the past have had a significant role in shaping the geological and biological history of the Earth [2][3]. The NEAs with Earth Minimum Orbit Intersection Distance (MOID) ≤ 0.05 AU and estimated diameter ≥ 150 m are classified as Potentially Hazardous Asteroid (PHA) and pose a potential

---

[*] Ph.D. Candidate, School of Aerospace Engineering; yumail2011@163.com

[†] Associate Professor, School of Aerospace Engineering; gongsp@tsinghua.edu.cn (Corresponding Author)

threat to the Earth[*]. In recent years, more and more space missions have begun to focus on the exploration of NEAs, such as the Near-Earth Asteroid Rendezvous (NEAR)-Shoemaker [4], OSIRIS-Rex [5], Hayabusa [6], and Hayabusa 2 [7]. Since the beginning of the introduction of the concept of solar sailing in 1920s [8], it has been proved of significant advantages in the field of deep space exploration. Because of the feature of propellant-free flight, solar sailing is one of the most promising tools for NEA exploration, such as impacting the asteroid with a high relative velocity for mitigating impact threat to the Earth [9], offering wide launch windows for rapid NEA rendezvous missions [10], and long-term multi-NEA rendezvous and sample return missions [11][12][13].

To reduce the cost of a single launch and increase the benefit of the spacecraft for NEA exploration, multiple targets exploration missions have significant roles. For such consideration, the Global Trajectory Optimization Competition (GTOC) has shown great interest in the mission design of multi-asteroid exploration, and over the 9 competitions organized to date, 4 were multi-asteroid rendezvous problems [14]. For example, the sponsor of GTOC 4 focused on the achievement of visiting as many NEAs as possible within a given time duration [15]. When considering a multi-target exploration mission, the preliminary design of trajectory becomes challenging because of the complexity of combinatorial problem that emerges with the increasing number of targets to visit in the mission [14]. On the one hand, the cost of transfer between two adjacent candidate targets, such as fuel consumption or transfer time, usually requires to solve a complicated numerical optimization problem, which is particularly of low computational efficiency in the preliminary mission design [16]. On the other hand, the complexity of global sequence search increases factorially as the search space increases, and it will cause catastrophic computing difficulty when facing a vast search space. To overcome those problems, different methods have been proposed in the past years. To the best knowledge of authors, the general ideas of those methods are to estimate the consumption of orbital transfer by sorts of approximate simplified models, such as impulsive model[17] and shape-based approach[18], and search the sequence using kinds of tree search method, such as branch-and-prune[19] and beam search[20]. For example, in GTOC 5, the team of Tsinghua University combined a ballistic approximation method and bounded and pruned algorithms in the roughly global search phase[17]; and the team from European Space Agency and University of

---

[*] NEO Basics – Potentially Hazardous Asteroids (PHAs), CNEOS NASA/JPL, https://cneos.jpl.nasa.gov/about/neo_groups.html [Retrieved 22 July 2018].

Florence proposed a linearized model of the "self-fly-by" to put a quick estimation of the propellant consumption and flight time for the transfer leg, which aided a first broad tree search of chemical propulsion options [19]. Considering a multi-NEA rendezvous mission for solar sails, Peloni et al. presented a shape-based approach to approximate the trajectories and a search-and-prune algorithm for sequence search [18]. For the preliminary design of multi-target problem, Izzo et al. proposed sorts of state-of-art methods, such as machine learning, and increase the efficiency and accuracy of estimating the fuel consumption of low-thrust transfer [21][22][23].

Recently, the Artificial Intelligence (AI) technologies has been successfully applied in many fields and dramatically improved the state-of-the-art in image and speech recognition, object detection and many other domains such as drug discovery and genomics [24]. The application of such techniques in the field of aerospace is getting more attention as well. Over the last few years, researchers have shown the possibility of using deep neural networks (DNN), one of the most promising AI technologies, for onboard real-time representation of optimal control profile for both landing problems and interplanetary trajectory transfer problems [25][26], parameter identification for detection and characterization of aircraft icing [27], and trajectory optimization of unmanned aerial vehicle (UAV) [28]. In these applications, the DNN achieves satisfactory accuracy and improves the efficiency greatly meanwhile. Inspired by the literature, we attempt to explore the possibility of using DNN for fast estimation of the transfer time for solar sails in the interplanetary trajectory.

Monte Carlo Tree Search (MCTS) is a method for searching optimal decision by building an incremental and asymmetric tree [29]. It has received considerable interest in game AI and combinatorial optimization problems, and the most well-known case is its successful application in AlphaGo [30][31]. The MCTS method iteratively repeats the selection, expansion, simulation and backpropagation process, and gives current optimal solution when terminated at any time or other stopping criteria. Compared to the traditional tree search methods, such as branch-and-bound and beam search, the MCTS method can give a better performance with limited computing resources, especially when facing planning problem with complex and colossal decision space, such as Go [29]. Because of the excellent performance of MCTS, it has been extended and applied in more regions, such as Traveling Salesman Problem (TSP) and interplanetary trajectory planning [32][33]. When used for the preliminary design of the interplanetary trajectory, the MCTS is able to find the best result that very close to the one flown by the real mission. In order to expand the application scope of this method, and to perform sequence search under limited computing resources, the MCTS method will be adopted in this work.

In this paper, we will first propose the state-of-art method to estimate the transfer time for solar sails in the interplanetary transfer trajectories using DNN. Then, based on the DNN model, we will use the MCTS method to search the exploration sequence among a NEAs set. To expand the practical scope of the MCTS method, a list of asteroids that of significant scientific value is pre-selected for further combinatorial optimization to search an optimal sequence that consumes the shortest mission time. Finally, by solving the subsequence leg by leg as an optimal control problem, the sequences with estimated transfer times will be validated.

This paper is organized as follows: In Section II, the dynamics model used in our work and the traditional method solving the optimal control problem to generate training database and the trajectory optimization is introduced. Section III presents the details of building the DNN model for estimating the transfer time in the interplanetary transfer trajectory. In Section IV, we will describe the MCTS algorithm and the sequence search process. Section V presents the sequence search results and verification of the searched sequences, and the results of the method described previously will be discussed. Finally, Section VI leads to our conclusions.

## II. Dynamic Model and Optimization Problem

### A. Solar-Sail Dynamic Model

The planar and perfectly reflecting model of the solar sail is assumed in this paper. As shown in Fig. 1, we use two angles, cone angle $\alpha$ and clock angle $\delta$, to describe the orientation of the solar sail. The cone angle $\alpha$ is defined as the angle between the sail surface normal $\hat{\boldsymbol{n}}$ and the incident radiation direction $\hat{\boldsymbol{r}}$, and the clock angle $\delta$ is defined as the angle between the solar sail heliocentric angular momentum $\hat{\boldsymbol{h}}$ and the projection of $\hat{\boldsymbol{t}}$ onto the plane normal to the radial direction of solar sail. The acceleration of the solar sail at a distance $r$ from the sun can be written as:

$$\boldsymbol{a}_s = \beta \frac{\mu}{r^2} \cos^2 \alpha \hat{\boldsymbol{n}} \tag{1}$$

where $\beta$ is the lightness number, $\mu$ is the solar gravitation constant, and the sail surface normal can be written as [34]:

$$\hat{\boldsymbol{n}} = \cos\alpha\hat{\boldsymbol{r}} + \sin\alpha\cos\delta\hat{\boldsymbol{h}} + \sin\alpha\sin\delta\hat{\boldsymbol{t}} \qquad (2)$$

The definition domain of the two angles is as follows:

$$\alpha \in [0, \pi/2], \delta \in [0, 2\pi] \qquad (3)$$

Thus, the dynamic equations considering a two-body model can be written as:

$$\begin{cases} \dot{\boldsymbol{r}} = \boldsymbol{v} \\ \dot{\boldsymbol{v}} = -\dfrac{\mu}{r^3}\boldsymbol{r} + \boldsymbol{a}_s \end{cases} \qquad (4)$$

where $r$ and $v$ are the state parameters of the solar sail.

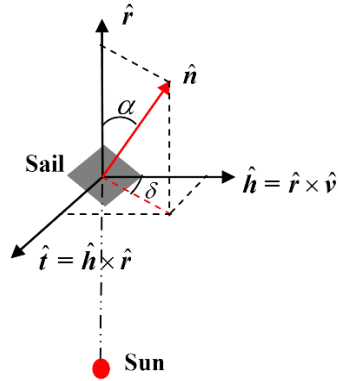

Fig. 1 Reference frame of the solar radiation pressure acceleration

For the convenience of calculation, the solar gravitation constant $\mu$ is normalized to 1 by normalizing the distance unit and time unit using the astronomical unit (AU) and $1/2\pi$ year, respectively[35][36]. Therefore, the dynamic equations after normalization can be written as

$$\begin{cases} \dot{\boldsymbol{R}} = \boldsymbol{V} \\ \dot{\boldsymbol{V}} = -\dfrac{1}{R^3}\boldsymbol{R} + \beta\dfrac{1}{R^2}\cos^2\alpha\hat{\boldsymbol{n}} \end{cases} \qquad (5)$$

where $R$ and $V$ are the normalized position and velocity vectors.

## B. Optimal Control Problem Formulation

Consider that a solar sail departs from the departure celestial body and rendezvous the arrival celestial body with free time, and the departure and arrival times are $t_0$ and $t_f$, respectively. Without the consideration of the fuel consumption, the objective function of trajectory optimization for the solar sail is the total time of flight (TOF) as follows

$$\min \ J = \int_{t_0}^{t_f} \lambda_0 dt \tag{6}$$

where $\lambda_0$ is a positive weight constant.

Assume that the initial and terminal states of the solar sail are the same as that of the departure and arrival celestial body, which defines the boundary condition of the optimal control problem. Thus, the state constraints can be described as follows

$$\boldsymbol{\Psi}(t_0) = \begin{cases} \boldsymbol{R}(t_0) - \boldsymbol{R}_d(t_0) \\ \boldsymbol{V}(t_0) - \boldsymbol{V}_d(t_0) \end{cases} = \boldsymbol{0} \tag{7}$$

$$\boldsymbol{\Psi}(t_f) = \begin{cases} \boldsymbol{R}(t_f) - \boldsymbol{R}_a(t_f) \\ \boldsymbol{V}(t_f) - \boldsymbol{V}_a(t_f) \end{cases} = \boldsymbol{0} \tag{8}$$

where $\boldsymbol{R}$ and $\boldsymbol{V}$ are the position and velocity of the solar sail, $\boldsymbol{R}_d$ and $\boldsymbol{V}_d$ are the position and velocity of the departure celestial body, and $\boldsymbol{V}_a$ and $\boldsymbol{V}_a$ are the position and velocity of the arrival celestial body, respectively.

To introduce the Hamiltonian function of the system, the co-state variables can be defined according to the state variables as $\boldsymbol{\lambda} \triangleq (\boldsymbol{\lambda_R}; \boldsymbol{\lambda_V})$. Therefore, the Hamiltonian function of the system can be given as

$$H = \lambda_0 + \boldsymbol{\lambda}_R \cdot \boldsymbol{V} + \boldsymbol{\lambda}_V \cdot \left( -\frac{\boldsymbol{R}}{R^3} + \beta \frac{1}{R^2} \cos^2 \alpha \hat{\boldsymbol{n}} \right) \tag{9}$$

We can obtain the Euler-Lagrange equations via the derivative of the state variable by the Hamilton function as

$$\begin{aligned} \dot{\boldsymbol{\lambda}}_R &= -\frac{\partial H}{\partial \boldsymbol{R}} = \frac{1}{R^3} \boldsymbol{\lambda}_V - \frac{3}{R^5}(\boldsymbol{R} \cdot \boldsymbol{\lambda}_V)\boldsymbol{R} - 2\beta \frac{\cos \alpha}{R^3}(\boldsymbol{\lambda}_V \cdot \hat{\boldsymbol{n}})(\hat{\boldsymbol{n}} - 2\frac{\cos \alpha}{R}\boldsymbol{R}) \\ \dot{\boldsymbol{\lambda}}_V &= -\frac{\partial H}{\partial \boldsymbol{V}} = -\boldsymbol{\lambda}_R \end{aligned} \tag{10}$$

According to the optimal control theory, the boundary conditions satisfy the transversality conditions as

$$\lambda_R(t_0) = -\frac{\partial^T \Psi(t_0)}{\partial R(t_0)} \gamma_{R_0} = -\gamma_{R_0}$$

$$\lambda_V(t_0) = -\frac{\partial^T \Psi(t_0)}{\partial V(t_0)} \gamma_{V_0} = -\gamma_{V_0}$$

$$\lambda_R(t_f) = -\frac{\partial^T \Psi(t_f)}{\partial R(t_f)} \gamma_{R_f} = \gamma_{R_f} \quad (11)$$

$$\lambda_V(t_f) = -\frac{\partial^T \Psi(t_f)}{\partial V(t_f)} \gamma_{V_f} = \gamma_{V_f}$$

where $\gamma_0 \triangleq (\gamma_{R_0}; \gamma_{V_0})$ and $\gamma_f \triangleq (\gamma_{R_f}; \gamma_{V_f})$ are Lagrange multipliers and related to the initial and terminal co-state vectors.

In addition, the initial and terminal Hamiltonian functions are determined by the stationarity condition

$$H(t_0) = \gamma_0 \cdot \frac{\partial \Psi(t_0)}{\partial t_0} = \lambda_R(t_0) \cdot V_d(t_0) - \lambda_V(t_0) \cdot \frac{R_d(t_0)}{R_d^3(t_0)}$$

$$H(t_f) = \gamma_f \cdot \frac{\partial \Psi(t_f)}{\partial t_f} = \lambda_R(t_f) \cdot V_a(t_f) - \lambda_V(t_f) \cdot \frac{R_a(t_f)}{R_a^3(t_f)} \quad (12)$$

The optimal control law can be obtained by the Pontryagin's maximum principle as

$$\alpha^* = \begin{cases} \tan^{-1}\left(\dfrac{3+\sqrt{9+8\tan^2 \tilde{\alpha}}}{4\tan \tilde{\alpha}}\right) & \tilde{\alpha} < 90° \\ \tan^{-1}\left(\dfrac{3-\sqrt{9+8\tan^2 \tilde{\alpha}}}{4\tan \tilde{\alpha}}\right) & \tilde{\alpha} > 90° \end{cases} \quad (13)$$

$$\delta^* = 180° + \tilde{\delta}$$

where $\alpha^*$ and $\delta^*$ are the optimal control angles of solar sail, $\tilde{\alpha}$ and $\tilde{\delta}$ are the angles that describe the direction of the velocity co-state as follows[34][36].

$$\hat{\lambda}_V = \cos\tilde{\alpha}\hat{r} + \sin\tilde{\alpha}\cos\tilde{\delta}\hat{h} + \sin\tilde{\alpha}\sin\tilde{\delta}\hat{t} \quad (14)$$

## C. The Process of Solving Two-Point Boundary Value Problem

Based on the formulation introduced in the previous subsection, the indirect method to numerically solve the optimal control problem will be adopted. Generally, the indirect method solves the problem by converting the original optimal control problem to a boundary-value problem, and the optimal solution is found by solving a system of differential equations that satisfy the boundary conditions [37]. Before we solve the two-point boundary value problem, the normalization of the initial co-state vector is performed to improve the efficiency of indirect shooting method. As proposed by Jiang [38], the co-state vectors can be normalized as

$$\lambda \triangleq \frac{\lambda}{\|\lambda(t_0)\|} = \frac{(\lambda_0; \lambda_R; \lambda_V)}{\|\lambda_0; \lambda_R(t_0); \lambda_V(t_0)\|} \quad (15)$$

where $\lambda_0$ is kept as constant, and the initial co-state vector satisfies the constraints as Eq.(16).

$$\|\lambda(t_0)\| = \sqrt{\lambda_0 + \|\lambda_R(t_0)\| + \|\lambda_V(t_0)\|} = 1 \quad (16)$$

Consequently, the six-dimensional unbounded co-state vector can be bounded on the seven-dimensional unit sphere, which reduces the difficulty of guessing the initial co-states significantly.

$$\boldsymbol{X}_0 = \left[ t_0;\ t_f;\ \lambda_0;\ \lambda_R;\ \lambda_V \right] \quad (17)$$

$$\begin{cases} f_{H_0}(t_0) = H(t_0) - \gamma_0 \cdot \dfrac{\partial \boldsymbol{\Psi}(t_0)}{\partial t_0} \\[6pt] f_{H_f}(t_f) = H(t_f) - \gamma_f \cdot \dfrac{\partial \boldsymbol{\Psi}(t_f)}{\partial t_f} \\[6pt] f_{\lambda_0}(\lambda_0) = \sqrt{\lambda_0 + \|\lambda_R(t_0)\| + \|\lambda_V(t_0)\|} - 1 \\[6pt] \boldsymbol{\Psi}(t_f) = \begin{Vmatrix} \boldsymbol{r}(t_f) - \boldsymbol{R}_a(t_f) \\ \boldsymbol{v}(t_f) - \boldsymbol{V}_a(t_f) \end{Vmatrix} \end{cases} \quad (18)$$

Therefore, as shown by Eq.(17), there are nine initial values to be determined when considering the rendezvous problem with a free departure time, including the departure and arrival time and seven-dimensional values of the co-states. If the departure time of the rendezvous problem is fixed, initial values to be determined will be reduced to eight, and the first equation in Eq.(12) will be ignored consequently. The roots of the nonlinear function as Eq.(18) will be searched by shooting method and the MinPack-1, a package for the numerical solution of systems of nonlinear equations and nonlinear least square problems, will be used [39]. In each case, the initial guess will be randomly generated for 1000 times to solve the nonlinear equations until the residuals as Eq.(18) meet the error requirements.

## III. Transfer Time Mapping Using DNN

As a mapper for predicting the orbital transition time, the input of DNN is the orbital feature of the departure and arrival celestial bodies, and the output is the corresponding orbital transition time, as shown in Fig. 2. In this section, we will introduce the process of building the DNN model for mapping the transfer time for solar sails. First, the method of generating training and validation sample set is described.

Different number of hidden layer and unit in each layer, as well as different activation functions, are considered for parameter tuning of the network. At last, two test sets are used to evaluate the trained DNN model.

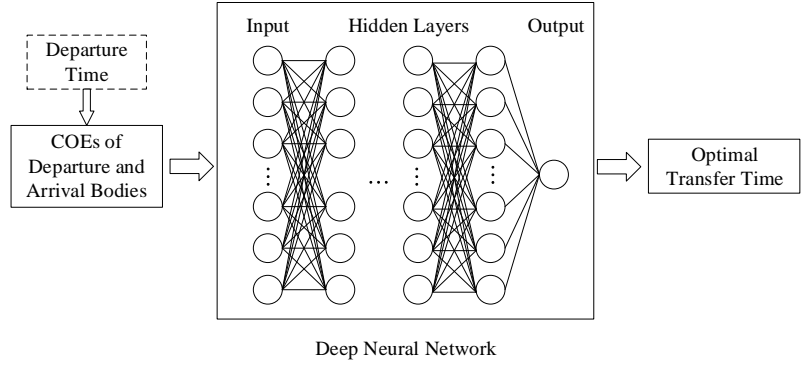

**Fig. 2 Mapping relationship of the DNN**

## A. Generation of Training and Validation Database

Consider the rendezvous problem formulation in the previous section. The aim of the DNN model is to establish a mapping relationship from the given boundary conditions to the optimal orbital transfer time. Considering the multi-asteroid sequence planning problem in the following work, the fixed departure time rendezvous problem is assumed. Suppose that the solar sail leaves the departure celestial body at Modified Julian Date (MJD) 57800.00, and the lightness number of the solar sail is kept constant as $\beta$=0.1265 (corresponding to the characteristic acceleration as $a_c$=0.75mm/s$^2$). To ensure the universality of sample set, the classical orbital elements (COEs) of the departure and arrival celestial bodies at this moment are randomly generated around the Earth's orbit. Consider the rendezvous problem between the NEAs, as shown in Fig. 3, 40000 of pairs of random COEs are generated as Eq.(19).

$$\begin{aligned}
a &= 1.0 + 0.2 \times rand[-1,1] \, (AU) \\
e &= 0.0 + 0.2 \times rand[0,1] \\
i &= 0.0 + 0.2 \times rand[0,1] \, (rad) \\
\Omega &= 0.0 + 2\pi \times rand[0,1] \, (rad) \\
\omega &= 0.0 + 2\pi \times rand[0,1] \, (rad) \\
f &= 0.0 + 2\pi \times rand[0,1] \, (rad)
\end{aligned} \quad (19)$$

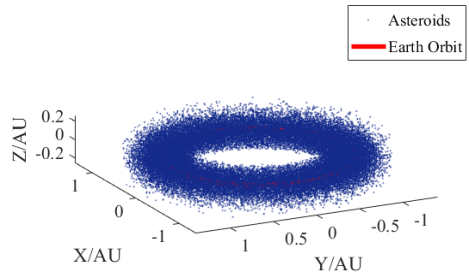

**Fig. 3 Visualization of training database for deep learning**

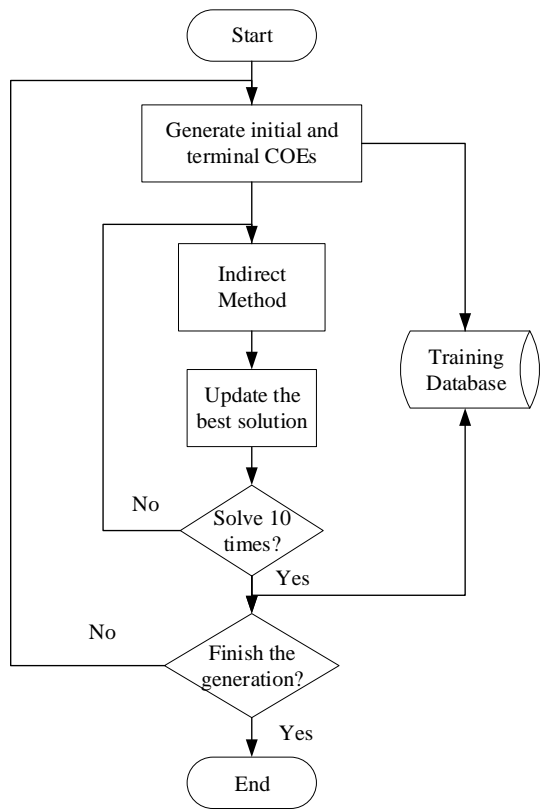

**Fig. 4 Process of the generation of the database**

These generated COE samples do not actually represent the ephemeris of real NEAs, but to some extent conform to the orbital characteristics of NEAs. Therefore, we can use these *pseudo-COEs* to calculate the orbital transfer time between them to train the DNN model. What we expect is to use these COE samples to obtain a DNN model that accurately predicting the orbital transition time between two arbitrary real NEAs. For each pair of COEs, the indirect method used to solve the time-optimal control problem will be executed for ten times and the best solution with shortest transfer time is stored as the optimal result. The process of generating sample set is shown in Fig. 4. We can obtain 40000 sets of COEs at the departure time and the corresponding optimal transfer time. Based on these sample sets, the mapping relationship from the COEs to the optimal transfer time will be established by DNN.

**B. Parameter Tuning of DNN**

Consider the fully-connected DNN shown as Fig. 2. Before we build the network, there are several aspects to be considered. The first is the feature description method of input data. In addition to the COE description as mentioned previously, to achieve better mapping effect, the description method of position and velocity (RV) and modified orbital element (MOE) are also worth considering. In addition, in the absence of sufficient experience with different DNN structures and the corresponding effects, parameter tuning of the network is necessary. The architecture of the network to be determined includes the number of the hidden layer and unit in each layer, the activation function of hidden layers, and the training algorithm. Other hyper-parameters, such as the batch size and initial learning rate, are tuned according to the performance of neural networks and the strategies introduced in the literature [40][42]. For example, the optimal learning rate is usually set close to the largest that does not cause divergence of the training criterion. In our research, the learning rate is firstly set as $\eta = 0.01$, and if the training criterion shows apparent divergence, another value will be tried, such as a value three times smaller [40]. The batch size is tuned from 10 to 500, and the size of 200 is adopted for most cases. When the training set is wholly visited batch by batch, that is to say, when a training epoch is completed, the data in the training set will be randomly disrupted to ensure the generalization of sample selection in each training epoch.

The loss function in the form of mean square error (MSE) and the accuracy function in the form of mean absolute percentage error (MAPE) are defined as

$$\xi_{loss} = \sum_{i=1}^{n} \frac{\left(y_i - f(x_i)\right)^2}{n} \qquad (20)$$

$$\xi_{accuracy} = 1 - \sum_{i=1}^{n} \frac{\left|y_i - f(x_i)\right|}{n \cdot y_i} \qquad (21)$$

where $n$ is the size of the training or validation set at each training iteration, $y_i$ is the true label value, and $f(x_i)$ is the output of the network. Meanwhile, the mean absolute error (MAE) between the predicted value and true value is also used to measure the prediction effect of the model [41]:

$$\varepsilon_{MAE} = \frac{1}{n} \sum_{i=1}^{n} \left|y_i - f(x_i)\right| \qquad (22)$$

We divide the previously generated samples into two parts, 90% and 10%, as the training set and validation set, respectively. To test the effect of different data feature descriptions on training DNN, a network with four hidden layers and 60 units in each layer is built. The sigmoid function is adopted as the activation function of hidden layers. After being trained using the same training set for 10000 epochs and tested using the same test set, the fitting effect of different data feature description methods are compared, as shown in Tab. 1.

Tab. 1 Fitting effect of different data feature descriptions

| Feature Description | Test Accuracy (%) | Test Accuracy Less Than 300 days (%) | MAE (day) |
|---|---|---|---|
| RV | 92.602 | 87.588 | 25.843 |
| **COE** | **97.159** | **92.530** | **14.308** |
| MOE | 93.026 | 85.838 | 23.974 |

As shown in Tab. 1, the feature description using COE has the best fitting accuracy and the least MAE, and leads to the best prediction performance. It is worth mentioning that, in the following multi-asteroid sequence planning process, the output with shorter transfer time will be of more interest, so the fitting effect of the results less than 300 days are concerned and compared separately. In the following DNN parameter tuning and testing, we will use *local accuracy* to represent the fitting effect of the transfer time outputs below 300 days.

The activation function in hidden layers can also affect the training and fitting effect of DNN. We consider three commonly used activation functions, *ReLu*, *sigmoid* and *tanh*, and compare their performance in the training process. After training a DNN with four hidden layers and 60 units for 10000 epochs, the results are listed in Tab. 2. For different activation functions, the final fitting effects and training speeds are different as well. Among the three activation functions, the ***sigmoid*** function has the fastest convergence speed and the best fitting results.

Tab. 2 Performance of different activation functions

| Activation Function | Test Accuracy (%) | Local Accuracy (%) | MAE (day) |
|---|---|---|---|
| tanh | 95.449 | 82.227 | 26.594 |
| **sigmoid** | **97.159** | **92.530** | **14.308** |
| ReLu | 94.126 | 79.498 | 32.222 |

What we are mostly concerned with is the structure of the networkt, including the number of the hidden layer and unit in each layer. If the size of the network is too small, it can not reflect the mapping relationship very well. Otherwise, it will restrain the convergence speed of the training and leads to worse fitting effect because of inadequate training. To achieve a reasonable tradeoff between fitting effect and convergence performance, the number of the hidden layer from 1 to 7 and unit in each layer from 12 to 240 are considered. At first, we consider the neural network with three hidden layers and different number of units in each hidden layer. Then, the neural network with different number of hidden layers and 108 units in each hidden layer are trained and recorded. The results are shown in Tab. 3, Tab. 4, and Fig. 5.

**Tab. 3 Fitting results of DNN with 3 hidden layers and different number of unit**

| Number of Unit | Test Accuracy (%) | Local Accuracy (%) | MAE (day) |
|---|---|---|---|
| 12 | 92.921 | 75.730 | 27.892 |
| 24 | 95.444 | 82.860 | 19.573 |
| 36 | 95.561 | 89.458 | 14.308 |
| 48 | 95.795 | 90.143 | 13.258 |
| 60 | 97.030 | 90.136 | 12.532 |
| 72 | 97.110 | 90.572 | 11.582 |
| 84 | 96.597 | 90.540 | 11.541 |
| 96 | 96.351 | 91.807 | 11.292 |
| 108 | 96.306 | 92.025 | 10.999 |
| **120** | **96.507** | **91.906** | **10.609** |
| 150 | 96.668 | 91.094 | 11.418 |
| 180 | 96.730 | 90.071 | 11.885 |
| 210 | 96.430 | 89.750 | 14.254 |
| 240 | 95.193 | 87.122 | 16.127 |

**Tab. 4 Fitting results of DNN with different number of hidden layer and 108 units**

| Number of Hidden Layer | Test Accuracy (%) | Local Accuracy (%) | MAE (day) |
|---|---|---|---|
| 1 | 93.576 | 80.541 | 23.939 |
| 2 | 96.217 | 89.735 | 18.381 |
| 3 | 96.306 | 92.025 | 10.999 |
| 4 | 97.159 | 92.530 | 11.150 |
| **5** | **97.134** | **92.954** | **10.176** |
| 6 | 96.924 | 92.236 | 10.378 |
| 7 | 96.824 | 92.120 | 12.013 |

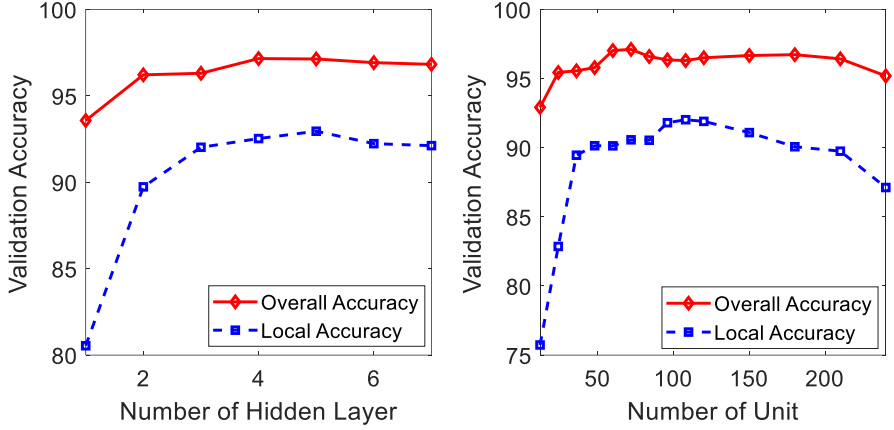

**Fig. 5 Fitting results of different layers and units**

According to the results, a neural network with shallow hidden layer or few units in each hidden layer cannot achieve a satisfactory fitting result. As the number of hidden layers increases, the fitting results of the neural network gets better, but the training time will be increased significantly at the same time. Additionally, with the increase of network structure, training demand for sample size and training epoch will increase as well. Therefore, as shown in Tab. 3, Tab. 4, with the same sample size and training epoch, the fitting effect of the network will become worse when the size of the network is too large (e.g., over 5 hidden layers and 150 units in each layer). Thus, a network with 3 to 5 hidden layers and 96 to 150 units in each hidden layer is more reasonable considering the tradeoff between training cost and fitting effect.

In addition, we have also compared the effects of different training algorithms. As listed in Tab. 5, the gradient descent algorithm has the best fitting effect.

**Tab. 5 Performance of different activation functions**

| Training Optimizer | Test Accuracy (%) | Local Accuracy (%) | MAE (day) |
|---|---|---|---|
| **Gradient Descent** | **97.197** | **92.632** | **10.592** |
| RMS Prop | 96.453 | 90.688 | 11.853 |
| Adam | 95.666 | 89.989 | 15.294 |
| Adagrad | 92.148 | 82.618 | 20.356 |

## C. Training and Evaluation

Based on the performance of the parameter tuning, we finally adopt the architecture of the DNN as shown in Tab. 6. The training and validation set contains 40000 samples, and the COE description is used to describe the orbital features. We build the network with five hidden layers and 120 units in each hidden layer, and the sigmoid function is adopted as the activation function of hidden layers. To get better training results, an exponential decay learning rate is adopted, and the batch size for each training iteration is set to $B = 200$. The learning rate descends along with the training epoch as Eq.(23), which satisfies the exponential decay rule as shown in Fig. 6.

$$\eta = \eta_0 \cdot \kappa^{\frac{epoch}{200}} \tag{23}$$

where $\eta_0 = 0.01$ is the initial learning rate, $\kappa = 0.98$ is the decay rate.

**Tab. 6 Architecture of the deep neural network**

| Parameter | Value |
| --- | --- |
| Feature description of input data | COE |
| Hidden layer and units in each layer | 5 × 120 |
| Activation function of hidden layers | sigmoid |
| Batch size | 200 |
| Initial learning rate | 0.01 |
| Training Optimizer | Gradient Descent |
| Training epoch | 30000 |

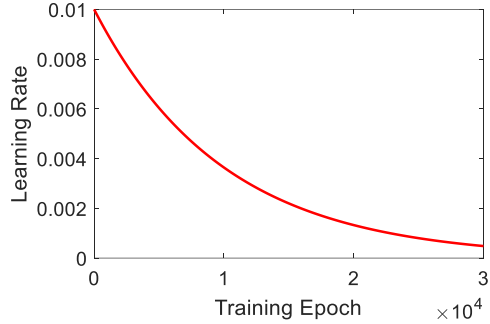

**Fig. 6 The learning rate decay in the training process**

The training results of DNN is shown in Tab. 7 and Fig. 7. The total fitting accuracy of the training and validation process is 98.352% and the local accuracy of the samples presenting the transfer time of less than 300 days is 94.441%. The loss value described as Eq.(20) is 0.174, and the MAE between the predicted and real orbital transfer time is 9.892 days.

**Tab. 7 Final fitting result of the deep neural network**

| Parameter | Value |
| --- | --- |
| Accuracy | 98.352% |
| Local Accuracy | 94.441% |
| MAE (day) | 9.892 |
| Final loss value | 0.174 |

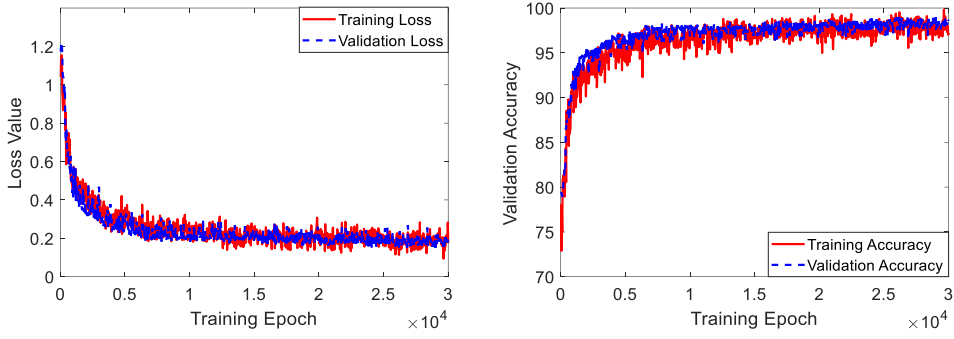

**Fig. 7 The loss and accuracy values in the training process**

To verify the trained DNN, tests are conducted with sample sets different from the training and validation set. As shown in Fig. 8a, the test set $A_1$ contains 100 pairs of test samples generated from the same domain as the training and validation set, and the fitting accuracy is up to 98.934%. Moreover, to study the behavior of the DNN, the test set $A_2$ shown as Fig. 8b is generated with the samples outside of the training set. In set $A_2$, the semi-major axis of the arrival body is generated randomly from 1.2 AU to 1.6 AU. The fitting accuracy tested with $A_2$ is 96.993%. The estimation error distribution of the transfer time is shown in Fig. 9. The estimation error for more than half of the test samples in test set $A_1$ are within 10 days, and the maximum error is 24 days, as shown in Fig. 9a. For test set $A_2$, the estimation errors for more than 80% test samples are less than 25 days, and the maximum error for 9 test samples exceeds 40 days, as shown in Fig. 9b.

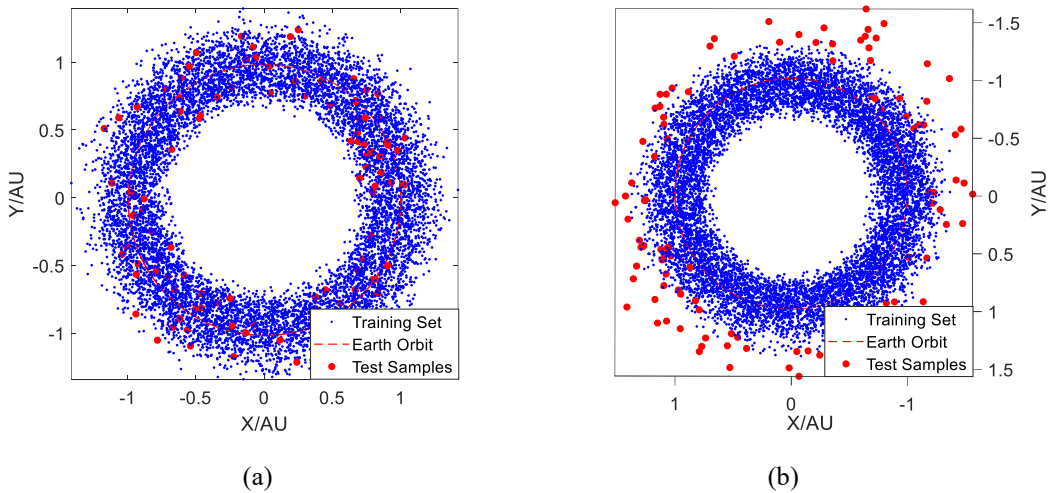

(a)          (b)

**Fig. 8 Test set distribution of $A_1$ (left) and $A_2$ (right)**

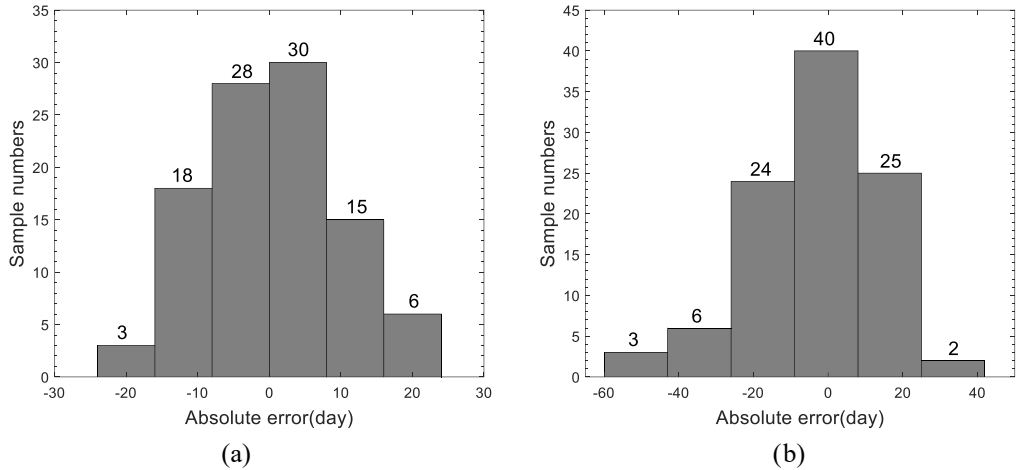

**Fig. 9 Absolute error distribution of test samples in $A_1$ (left) and $A_2$ (right)**

Furthermore, we also compare the effect of linear regression on estimating the transfer time, and the estimation error distribution is shown in Fig. 10. Using the same sample sets as training and testing DNN for linear regression, the fitting accuracy of the obtained regression function is only 83.355%. The MAE of the estimated transfer time is up to 69.333 days.

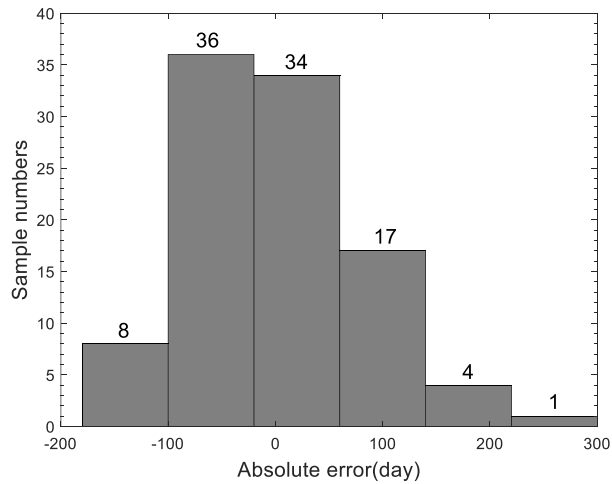

**Fig. 10 Estimation error distribution of linear regression**

According to the test results, the DNN can achieve an excellent mapping relationship from the orbital characteristics of the departure and arrival celestial bodies to the transfer time for solar sails. This means that the complicated nonlinear relationship from the orbital characteristics to orbital transfer time can be replaced by the DNN, which is of great significance in the sequence planning for multi-object missions using solar sail.

# IV. Sequence Search Using MCTS

In this section, we will present the algorithms and procedure for sequence search. Based on the DNN obtained previously, an MCTS method is adopted for target selection and sequence planning problem. We first introduce the primary process of the MCTS and Upper Confidence Bound for Tree (UCT) policy. Then, we will present the corresponding implementation methods for target selection and sequence planning problem. Considering different application backgrounds, different parameters of tree policy are tuned, and the performances are compared.

## A. The Process of Monte Carlo Tree Search

The MCTS method incrementally establishes an asymmetric search tree by randomly extracting samples in a given decision space. As shown in Fig. 11, the search procedure performs four processes repeatedly [29]. Each node in the search tree contains the information about the current state, executed choice, and the number of times visited to this node. For a given node, while all of the children of the current node have been fully expanded, the **_Selection_** process will be executed with a **_Tree Policy_** recursively until reaching a node that not all of the children are fully expanded, shown as Fig. 11a. Then, the **_Expansion_** process will be performed to randomly create a child node, namely a **_leaf node_**, shown as Fig. 11b. From the leaf node, the **_Simulation_** process will adopt the **_Default Policy_** to execute a random move strategy to get a fast estimation of the final reward, shown as Fig. 11c. Once the final reward is obtained, the **_Backpropagation_** process will go back through all the nodes upward to update their information, shown as Fig. 11 d.

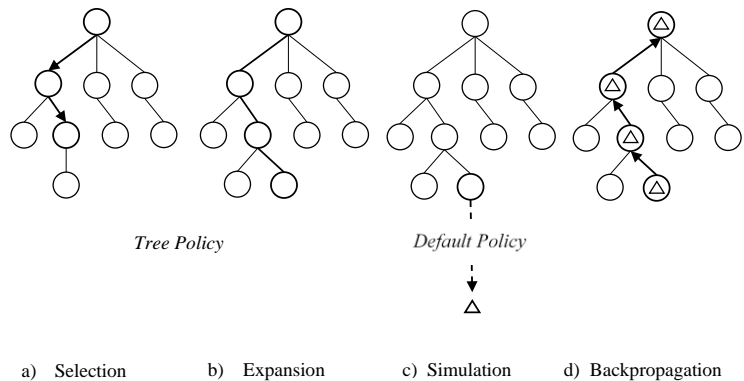

a) Selection    b) Expansion    c) Simulation    d) Backpropagation

**Fig. 11 Elemental composition of Monte Carlo Tree Search**

The general process of the MCTS is shown in Fig. 12. The four processes introduced above constitute one iteration step of MCTS. As the number of iterations increases, the complexity and required memory of the search tree grow. When reaching any desired computation time or iterations, the search process will be stopped, and the final decision will be made according to the tree policy.

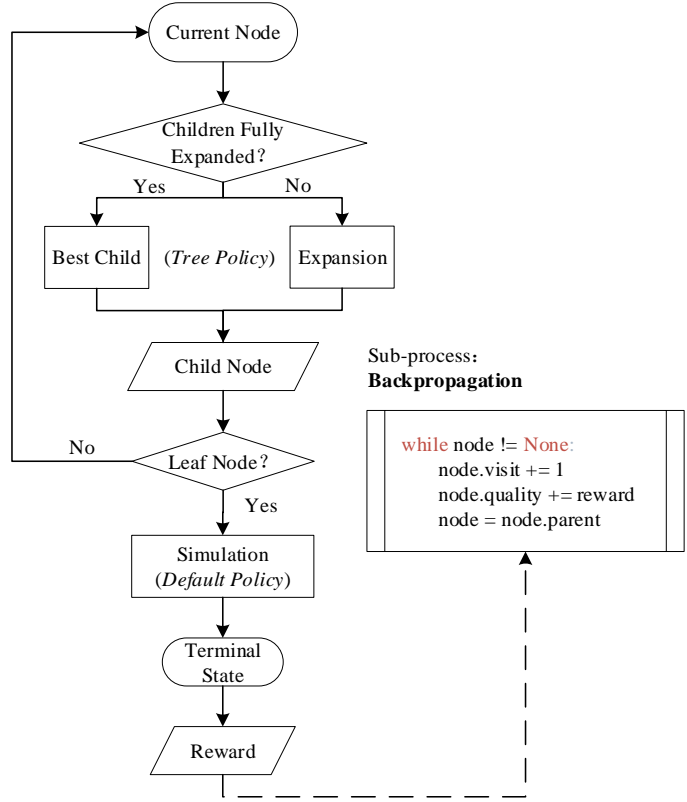

**Fig. 12 The general procedure of Monte Carlo Tree Search**

The tree policy plays a crucial role in the MCTS and affects the tradeoff between the exploration and exploitation. The most popular tree policy is the UCT algorithm proposed by Kocsis and Szepesvári [43]. The algorithm can be described as follows

$$\underset{i}{\arg\max}\ \frac{Q_i}{N_i} + c_p \sqrt{\frac{2\ln N}{N_i}} \tag{24}$$

where $N$ and $N_i$ represent the number of times the current node and its child node $i$ have been visited, $Q_i$ is the estimated reward for the child node $i$, and $c_p$ is a constant parameter which balances the left and right terms of this equation. When considering sequence planning problem, the tuning of the parameter $c_p$ affects the tradeoff between the exploration and exploitation. The parameter tuning and the corresponding effects will be presented in the rest this section.

## B. Sequence Search Strategy Using MCTS

The objective of trajectory optimization for solar sails is to minimize the TOF of the mission. Therefore, we propose two application backgrounds in the multi-asteroid rendezvous problem for MCTS method; one is sequence planning of some given targets to be visited, and the other one is target selection from a given asteroid set. These two applications require the search results to have the shortest flight time or to access as many targets as possible within the specified time. Based on those application backgrounds, the goal of MCTS is to search a sequence with the shortest flight time or a longest sequence within a specific TOF. When considering target selection problem, the maximum depth of the search tree, namely the longest length of the sequence, can be set to a small value at first, and if the TOF does not exceed the specified time, the maximum depth of the search tree will be increased. Thus, the strategy of MCTS for searching the longest sequence within a specific flight time can be sequentially converted to that of searching the sequence with the shortest flight time.

Therefore, we consider the multi-asteroid rendezvous problem with a specific sequence length. Given a departure time from the Earth, the transfer time from the Earth to the first target can be estimated using the DNN that we obtained previously. For the next transfer leg from the current target to the next, the transfer time will be estimated according to the current updated orbital elements of the targets. Therefore, each node in the search tree is designed to keep track of the selected target, current flight time, number of times this target has been considered, and the average quality of the selected target. The nodes in the $k$th layer of the search tree represent the $k$th considered targets to be accessed, which means the maximum depth of the search tree is equal to the length of the sequence. The search is terminated when the number of rendezvous targets specified in the sequence length is completed, and the final TOF is the output of the selected sequence. As the UTC algorithm described as Eq.(24) requires the reward of the sequence result to be distributed in the interval [0, 1], we convert the TOF of $\Delta T$ to the reward as follows.

$$\Delta = \frac{\Delta T_{\max} - \Delta T}{\Delta T_{\max}} \tag{25}$$

where $\Delta T_{max}$ is the longest acceptable flight time and $\Delta$ is the final reward of the sequence.

The sequence search strategy using MCTS can be described as Algorithm 1. The search tree is built layer by layer, and the MCTS process is performed for a specified number of simulation times in each layer to generate the best child, namely a selected target for next rendezvous. To increase the performance of the method, the MCTS search process is periodically initialized and restarted, and all of the search results are stored for a final decision to be reached [29][44]. After being performed for serval times, the best result among all the stored results will be regarded as the optimal solution.

| **Algorithm 1**   Sequence Search Strategy |
| --- |
| Initialize the initial state $s_0$ with Earth and the departure time from Earth |
| Create root node $n_0$ with $s_0$ |
| **for** $i$ in *max_depth*: |
|     **function** MCTS($n_0$): |
|         **for** *sim_time* in *max_simulation_times*: |
|             $n_l$ ← **Tree Policy**($n_0$) |
|             $\Delta$ ← **Default Policy**($n_l$) |
|             **Backpropagation**($n_l$, $\Delta$) |
|         $n_i$ ← **Best Child**($n_0$) |
|         **return** $n_i$ |
|     $n_0$ ← $n_i$ |

## C. Parameter Tuning of Tree Policy

First, we consider the sequence planning problem. Assuming that we have selected a set of target asteroids, and the purpose of MCTS is to plan the sequence order to rendezvous with these targets. We suppose that the mission background is to continuously rendezvous 15 NEAs using a solar sail with the lightness number as $\beta$=0.1265. When the solar sail reaches the target, it stays ten days with the asteroid for scientific exploration and then departs to the next asteroid. To get the best sequence, the parameter tuning of $c_p$ in Eq.(24) is performed.

As shown in Fig. 13, we compared the average, maximum, and minimum TOF of 100 sets of sequence results under each parameter. The average output represents the overall distribution of the results, and the maximum output indicates the maximum deviation between the results and the optimal solution. The minimum output represents the ability of the algorithm to reach the optimal solution.

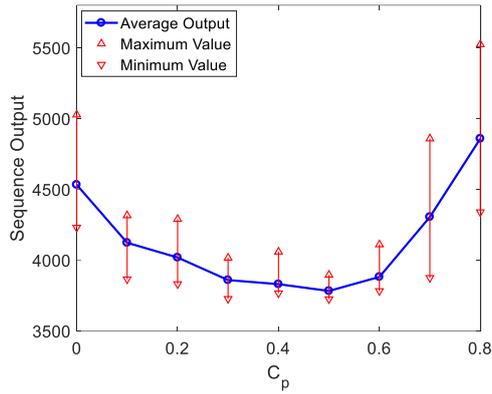

**Fig. 13 Parameter tuning of UCT algorithm for sequence planning**

The value of the parameter $c_p$ determines a tradeoff between left and right terms in Eq.(24). According to the distribution of the output shown in Fig. 13, when the value of $c_p$ is too large or small, which means the UCT policy considers either more exploration or more exploitation respectively, the output of the sequence is of poor performance. The values between 0.4 and 0.6 lead to a better performance no matter considering the average, minimum and maximum output.

Furthermore, we consider another application background of MCTS method, that is, target selection among a big asteroid set, which has the similar background as GTOC 5 [15]. Compared with the sequence planning problem, available choices of target selection problem in each step of the search tree are much more abundant, which implies that the previous target selection will not have much impact on the following target selection. We select a sequence of 15 targets from a set of 1000 NEAs. The performance of different values of the parameter $c_p$ is presented in Fig. 14. A depth-first tree search strategy of MCTS is necessary for the target selection applications. Thus the values of $c_p$ smaller than $1\times 10^{-3}$ are more suitable according to the parameter tuning result.

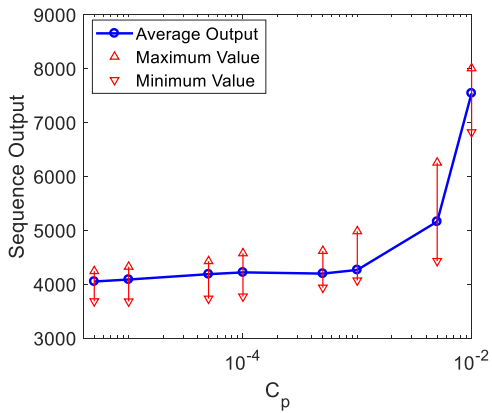

**Fig. 14 Parameter tuning of UCT algorithm for target selection**

# V. Sequence Search and Verification

In this section, the sequence search strategy introduced in the previous section will be performed by considering two different examples. The first example is required to select a sequence that can rendezvous as many targets as possible within the specified mission time. The other example is a sequence planning problem for several given NEAs. After the preliminary sequence search, method for solving the optimal control problem introduced in Section II will be used for local optimization and validation of the results obtained by preliminary design. The introduced MCTS method and the DNN model for estimating the transfer time of solar sails will be verified via two examples.

## A. Target Selection Problem

Suppose that the mission requires the solar sail to be launched from the Earth at MJD 64329 (2035-01-01) to rendezvous as many as possible NEAs within the TOF of 10 years. The lightness number of the solar sail is set as $\beta$=0.1265, and the accompanying flight time with each asteroid is 10 days. To narrow the target search space, an asteroid set that contains 1000 NEAs and satisfies the constraint presented as Eq.(19) is pre-selected from the asteroid library provided by JPL[*]. The number of simulation times of each child's selection in the MCTS process is set to 2000. To search a sequence with the longest length of asteroids, we perform the strategy introduced in Section IV and execute the search process for 200 times, and the best solution will be stored as the optimal solution.

The results as shown in Fig. 15 present the flight time of each transfer leg and the time of the solar sail to rendezvous with each asteroid. Based on the rendezvous sequence, the optimal control problem in each leg is solved successively, and the comparison with the results of the preliminary design are shown in Fig. 15 and Tab. 8.

Fig. 15a shows the transfer time predicted by DNN and the true time by solving the optimal control problem in each transfer leg, and Fig. 15b shows the rendezvous time with each asteroid of the prediction value of preliminary design and the true value after verification by solving the optimal control problems. As the results shown in Tab. 8, the TOF predicted by DNN in the preliminary design is 3552.861 days. After solving the optimal control problem leg by leg using the method described in Section II, the verification result of the TOF is 3541.545 days. The final deviation of the TOF is 10.316 days. The predictions of the transfer time by DNN are consistent with the actual value of optimization results, which verifies the accuracy of the DNN model we obtained in Section III.

---

[*] Data available online at https://ssd.jpl.nasa.gov/dat/ELEMENTS.NUMBR (ELEMENTS.UNNUMBR) [retrieved 22 February 2017].

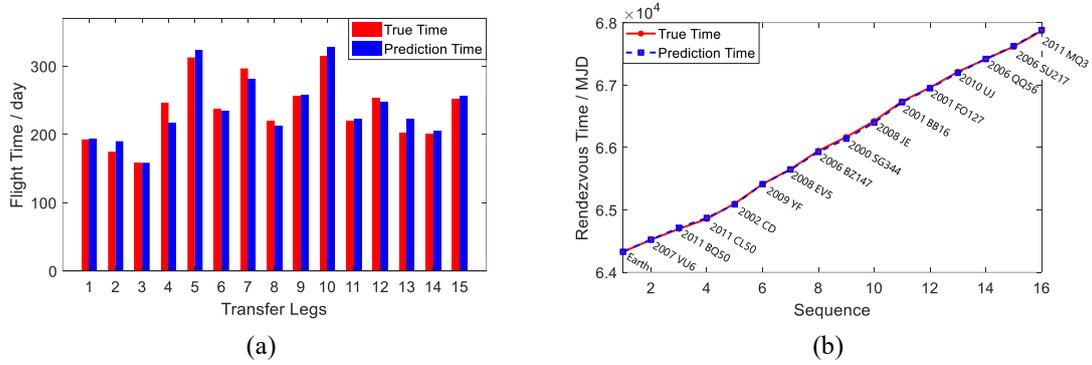

(a)      (b)

**Fig. 15 Transfer time and rendezvous time in each leg of the target selection problem**

**Tab. 8 Sequence search and verification result of target selection problem**

| Object | Rendezvous Time (MM-DD-YYYY) | Departure Time (MM-DD-YYYY) | Deviation From Validation Result (Day) |
|---|---|---|---|
| Earth | - | 01-01-2035 (01-01-2035) | 0 |
| 2007 VU6 | 07-13-2035 (07-05-2035) | 07-23-2035 (07-15-2035) | -8.468 |
| 2011 BQ50 | 01-04-2036 (01-11-2036) | 01-14-2036 (01-21-2036) | +6.873 |
| 2011 CL50 | 06-10-2036 (06-17-2036) | 06-20-2036 (06-27-2036) | +6.935 |
| 2002 CD | 02-12-2037 (01-31-2037) | 02-22-2037 (02-10-2037) | -12.334 |
| 2009 YF | 12-22-2037 (12-01-2037) | 01-01-2038 (12-11-2037) | -21.207 |
| 2008 EV5 | 08-17-2038 (08-02-2038) | 08-27-2038 (08-12-2038) | -14.534 |
| 2006 BZ147 | 06-10-2039 (05-21-2039) | 06-20-2039 (05-31-2039) | -19.463 |
| 2000 SG344 | 01-16-2040 (12-19-2039) | 01-26-2040 (12-29-2039) | -27.463 |
| 2008 JE | 09-28-2040 (09-02-2040) | 10-07-2040 (09-12-2040) | -26.190 |
| 2001 BB16 | 08-09-2041 (07-26-2041) | 08-19-2041 (08-05-2041) | -13.750 |
| 2001 FO127 | 03-17-2042 (03-07-2042) | 03-27-2042 (03-17-2042) | -10.520 |
| 2010 UJ | 11-26-2042 (11-09-2042) | 12-06-2042 (11-19-2042) | -17.207 |
| 2006 QQ56 | 06-17-2043 (06-20-2043) | 06-27-2043 (06-20-2043) | 2.758 |
| 2006 SU217 | 01-03-2044 (01-11-2044) | 01-13-2044 (01-21-2044) | 7.061 |
| 2011 MQ3 | 09-12-2044 (09-22-2044) | - | 10.316 |

*The values in the bracket are the results of preliminary design.

## B. Sequence Planning Problem

For the example to be presented in this part, we will illustrate the effectiveness of the DNN model and the MCTS method aiming at the sequence planning problem. Assume that there are several asteroids selected according to some criteria and the mission requires the solar sail to rendezvous with all of them, which is essentially a sequence planning problem. Inspired by the method used in [18] for asteroid database selection, we select NEAs from the database of Near-Earth Object Human Space Flight Accessible Target Study (NHATS)*. According to the criteria in Tab. 9 and constraint as Eq.(19), an asteroid set containing 15 NEAs is pre-selected for sequence planning.

**Tab. 9 Criteria for selecting the target for sequence planning**

| Constraints | Value |
| --- | --- |
| Total mission $\Delta v$ requirement | $\leq$ 6km/s |
| Total mission duration | $\leq$ 450 days |
| The number of days spent at the NEA | $\geqslant$ 8 days |
| Launch window | 2035-2040 |
| Absolute visual magnitude in magnitude units | $\leq$ 26 |
| Orbit Condition Code (OCC) | $\leq$ 7 |

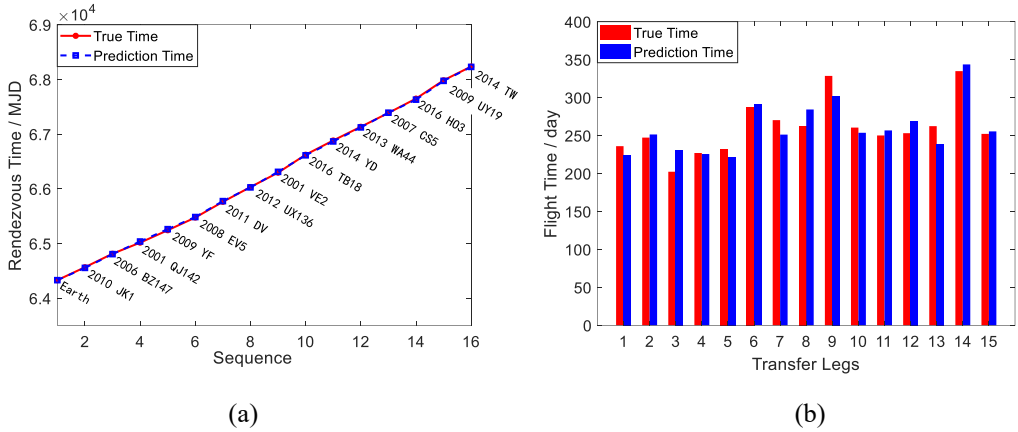

(a)  (b)

**Fig. 16 Transfer time and rendezvous time in each leg of the sequence problem**

The launch date, stay time with each asteroid and lightness number of the solar sail is the same as those of the previous example. The sequence planning strategy introduced in Section IV is used, and the parameter $c_p$ is set to 0.5. For 200 executions of the MCTS, the best visiting sequence requires the TOF of 3899.200 days, and this result appears for 114 times in the 200 executions. The average TOF is 4080.342 days, and the worst result leads to a TOF of 4618.995 days. The optimal sequence obtained by the MCTS method is listed in Tab. 10. The solar sail accomplishes its first rendezvous with 2010 JK1, and eventually rendezvous with 2014 TW to complete the mission. The local optimization results of the obtained optimal sequence are shown as in Fig. 16 and Tab. 10.

---

* Data available online at https://cneos.jpl.nasa.gov/nhats/ [retrieved 30 June 2018].

**Tab. 10 Sequence search and verification result of target selection problem**

| Object | Rendezvous Time (MM-DD-YYYY) | Departure Time (MM-DD-YYYY) | Deviation From Validation Result (Day) |
|---|---|---|---|
| **Earth** | - | 01-01-2035 (01-01-2035) | 0 |
| **2010 JK1** | 08-25-2035 (08-14-2035) | 09-04-2035 (08-24-2035) | -11.631 |
| **2006 BZ147** | 04-29-2036 (04-21-2036) | 05-09-2036 (05-01-2036) | -7.511 |
| **2001 QJ142** | 11-17-2036 (12-08-2036) | 11-27-2036 (12-18-2036) | +21.430 |
| **2009 YF** | 07-02-2037 (07-22-2037) | 07-12-2037 (08-01-2037) | +20.090 |
| **2008 EV5** | 02-19-2038 (03-01-2038) | 02-29-2038 (03-11-2038) | +9.589 |
| **2011 DV** | 12-04-2038 (12-17-2038) | 12-14-2038 (12-27-2038) | +12.829 |
| **2012 UX136** | 08-31-2039 (08-25-2039) | 09-10-2039 (09-04-2039) | -6.235 |
| **2001 VE2** | 05-20-2040 (06-04-2040) | 05-30-2040 (06-14-2040) | +15.283 |
| **2016 TB18** | 04-13-2041 (04-02-2041) | 04-23-2041 (04-12-2041) | -10.932 |
| **2014 YD** | 12-30-2041 (12-12-2041) | 01-09-2042 (12-22-2041) | -17.775 |
| **2013 WA44** | 09-06-2042 (08-25-2042) | 09-16-2042 (09-04-2042) | -11.132 |
| **2007 CS5** | 05-17-2043 (05-21-2043) | 05-27-2043 (05-31-2043) | +4.644 |
| **2016 HO3** | 02-03-2044 (01-15-2044) | 02-13-2044 (01-25-2044) | -19.191 |
| **2009 UY19** | 01-03-2045 (12-23-2044) | 01-13-2045 (01-02-2045) | -10.347 |
| **2014 TW** | 09-12-2045 (09-05-2045) | - | -7.060 |

*The values in the bracket are the results of preliminary design.

As shown in Fig. 16, the prediction and true values of verification coincide well, including the transfer time of each leg (Fig. 16a) and the rendezvous time with each asteroid in the sequence (Fig. 16b). The TOF of the prediction by DNN and verification result is 3899.200 days and 3906.260 days respectively, and the final deviation time of the mission duration is 7.060 days, as shown in Tab. 10.

## C. Discussion

From the examples above, the deviations of the TOF between the results of preliminary design based on the DNN model and the verified result are very small, which implies a relative error of less than 0.3%. At the same time, it is noteworthy that the predicted transfer time of each leg in the sequence does not have such a small error, such as the fourth leg in Fig. 15b and eighth leg in Fig. 16b. This difference is actually not a coincidence but rather determined by the characteristics of the orbit transfer window. When estimating the transfer time of each leg, a fixed departure time problem is considered. When the departure time moves, the TOF will be increased or decreased consequently. Therefore, when there is a small deviation of the estimation transfer time in the current transfer leg, the transfer time of the latter leg will also be increased or decreased by a small amount because of the change of the departure time in this leg, and will not have a large variation due to the slight movement of the departure window. For example, when considering the verification of the sequence as shown in Tab. 10 and Fig. 16, the predicted transfer time of the first leg is less than the result of verification, but because of the change of the departure time, the verified transfer time of the second leg is larger. According to the result in Tab. 10, the predicted rendezvous time with each asteroid varies around the time of verification, and the deviation will not be accumulated.

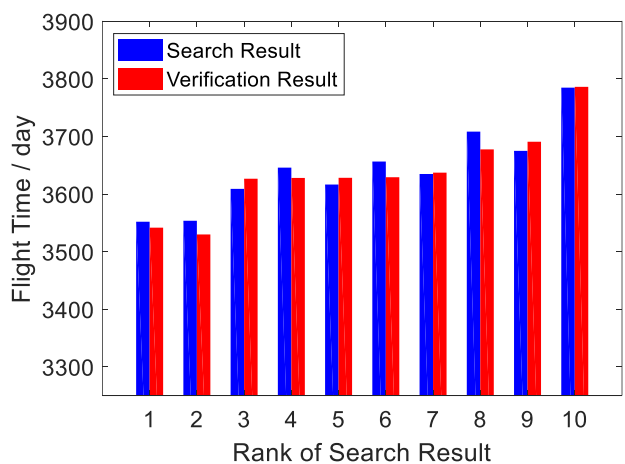

**Fig. 17 Predicted and true transfer time in the search result**

We also verified all the sequences that are in the top ten rank of the results. The predicted TOF and validation result for each sequence are shown in Fig. 17. The largest deviation in the results is the eighth sequence, which has a deviation of 31 days between the prediction and verification values and means a relative error of less 1.0%.

## VI. Conclusion

The multiple near-earth asteroids rendezvous problem is solved in this paper. The deep neural network has been established for mapping the relationship between the transfer time and the orbital characteristics. The classical orbital element is selected to describe the orbital characteristics. By the parameter tuning of the network, a deep neural network with 5 hidden layer and 120 units in each hidden layer is trained. According to the test results, the trained deep neural network achieves excellent mapping effectiveness from the orbital characteristics to the transfer time for solar sails. Furthermore, a test set is generated from the data outside of the training set, which results in the fitting accuracy of up to 96.993% with only a few results having large deviations. In the sequence search process of Monte Carlo tree search, two different applications are considered. For target selection problem and sequence planning problem, different parameters of the tree policy are investigated. The depth-first strategy is sufficient for the target selection problem, and a balanced exploration and exploitation policy has better performance for the sequence planning problem. The verification results of the searched sequence suggest that the proposed methods are effective.

## Acknowledgment

This work is supported by the National Natural Science Foundation of China (Grant No. 11772167 and 11822205).